\documentclass[10pt]{iopart}
\usepackage{graphicx}

\begin{document}

\title{Simulation of the White Dwarf --- White Dwarf galactic
  background in the LISA data.}

\author{Jeffrey A. Edlund\dag ,Massimo Tinto\dag \footnote[8]{Massimo.Tinto@jpl.nasa.gov},
        Andrzej Kr\'olak\ddag, and Gijs Nelemans\S}

\address{\dag\ Jet Propulsion Laboratory, California Institute of Technology,
               Pasadena, CA 91109}
\address{\ddag\ Max-Planck-Institute for Gravitational Physics,
                Albert Einstein Institute, D-14476 Golm, Germany
                \footnote[9]{On leave of absence from Institute of
              Mathematics,Polish Academy of Sciences, Warsaw, Poland}}
\address{\S\ Department of Astrophysics, IMAPP, Radboud University
             Nijmegen, The Netherlands}

%
\begin{abstract}
  LISA (Laser Interferometer Space Antenna) is a proposed space
  mission, which will use coherent laser beams exchanged between three
  remote spacecraft to detect and study low-frequency cosmic
  gravitational radiation. In the low-part of its frequency band, the
  LISA strain sensitivity will be dominated by the incoherent
  superposition of hundreds of millions of gravitational wave signals
  radiated by inspiraling white-dwarf binaries present in our own
  galaxy. In order to estimate the magnitude of the LISA response to
  this background, we have simulated a synthesized population that
  recently appeared in the literature. Our approach relies on entirely
  analytic expressions of the LISA Time-Delay Interferometric
  responses to the gravitational radiation emitted by such systems,
  which allows us to implement a computationally efficient and
  accurate simulation of the background in the LISA data. We find the
  amplitude of the galactic white-dwarf binary background in the LISA
  data to be modulated in time, reaching a minimum equal to about
  twice that of the LISA noise for a period of about two months around
  the time when the Sun-LISA direction is roughly oriented towards the
  Autumn equinox.  This suggests that, during this time period, LISA
  could search for other gravitational wave signals incoming from
  directions that are away from the galactic plane.  Since the
  galactic white-dwarfs background will be observed by LISA not as a
  stationary but rather as a cyclostationary random process with a
  period of one year, we summarize the theory of cyclostationary
  random processes and present the corresponding generalized spectral
  method needed to characterize such process. We find that, by
  measuring the generalized spectral components of the white-dwarf
  background, LISA will be able to infer properties of the
  distribution of the white-dwarfs binary systems present in our
  Galaxy.
\end{abstract}


\submitto{\CQG}

\maketitle

\section{Introduction}
\label{intro}

The Laser Interferometric Space Antenna (LISA) is a space mission
jointly proposed to the National Aeronautics and Space Administration
(NASA) and the European Space Agency (ESA). Its aim is to detect and
study gravitational waves (GW) in the millihertz frequency band. It
will use coherent laser beams exchanged between three identical
spacecraft forming a giant (almost) equilateral triangle of side $5
\times 10^6$ kilometers. By monitoring the relative phase changes of
the light beams exchanged between the spacecraft, it will extract the
information about the gravitational waves it will observe at
unprecedented sensitivities  \cite{PPA98}.

The astrophysical sources that LISA is expected to observe within its
operational frequency band ($10^{-4} - 1$ Hz) include extra-galactic
super-massive black-hole coalescing binaries, stochastic gravitational
wave background from the early universe, and galactic and
extra-galactic coalescing binary systems containing white dwarfs and
neutron stars.

Recent surveys have uniquely identified twenty binary systems emitting
gravitational radiation within the LISA band, while population studies
have concluded that the large number of binaries present in our own
galaxy should produce a stochastic background that will lie
significantly above the LISA instrumental noise in the low-part of its
frequency band.  It has been shown in the literature (see \cite{NYP01}
for a recent study and \cite{HBW90,EIS87} for earlier investigations)
that these sources will be dominated by detached white-dwarf ---
white-dwarf (WD-WD) binaries, with $1.1 \ \times 10^8$ of such systems
in our Galaxy.  By using the distribution for the parameters of the
WD-WD binaries given in (\cite{NYP01}) we have simulated the LISA
response to the WD-WD background.

This paper is organized as follows. In Section \ref{LISA_RES} we
recall the analytic expression of the unequal-arm Michelson
combination, $X$, of the LISA Time-Delay Interferometric (TDI)
response to a signal radiated by a binary system. In Section
\ref{WD_POP} we give a summary of how the WD-WD binary population was
obtained, and a description of our numerical simulation of the $X$
response to it. In Section \ref{NumSim} we describe the numerical
implementation of our simulation of the LISA $X$ response to the WD-WD
background, and summarize our results. The time-dependence and
periodicity of the magnitude of the WD-WD galactic background in the
LISA data implies that it is not a stationary but rather a
cyclostationary random process of period one year.  In Section
\ref{sec:cyclo} we provide a brief summary of the theory of
cyclostationary random processes relevant to the LISA detection of the
WD-WD galactic background, and apply it to three years worth of
simulated LISA $X$ data in Section \ref{ANALYSIS}. We find that, by
measuring the generalized spectral components of such cyclostationary
random process, LISA will be able to infer key-properties of the
distribution of the WD-WD binary systems present in our own Galaxy.

The non-stationarity of the WD-WD background was first pointed out by
Giampieri and Polnarev \cite{GP97} under the assumption of sources
distributed anisotropically, and they also obtained the Fourier
expansion of the sample variance and calculated the Fourier
coefficient for simplified WD-WD binary distributions in the Galactic
disc. What was however not realized in their work is that this
non-stationary random process is actually cyclostationary, i.e.  there
exists cyclic spectra that can in principle allow us to infer more
information about the WD-WD background than one could obtain by just
estimating the zero-order spectrum.

\section{The LISA response to signals from binary systems}
\label{LISA_RES}

The gravitational wave response $X^{\rm GW} (t)$ of the first generation
Michelson TDI combination to a signal from a binary system has been
derived in \cite{KTV04}, and it can be written in the following form
\begin{equation}
X^\mathrm{GW} (t) = Re \left[A(x, t) \ e^{-i \phi (t)}\right] \ ,
\label{X}
\end{equation}
where $x = \omega_s L$ ($\omega_s$ being the angular frequency of the
GW signal in the source reference frame), and the expressions for the
complex amplitude $A (x, t)$ and the real phase $\phi (t)$
are
\begin{eqnarray}
A(x, t) & = & 2 \, x \, \sin(x) \left\{ \left[ sinc[(1+c_2(t))\frac{x}{2}]
\ e^{i x(\frac{3}{2} + d_2 (t))} + sinc[(1-c_2(t))\frac{x}{2}] \ e^{i x(\frac{5}{2} +  d_2 (t))}
\right]
\ {\cal B}_2 (t) \right.
\nonumber
\\
& - & \left. \left[ sinc[(1-c_3(t))\frac{x}{2}] \ e^{i x(\frac{3}{2} +
  d_3 (t))}
+ sinc[(1+c_3(t))\frac{x}{2}] \ e^{i x(\frac{5}{2} +  d_3 (t))}
  \right] \ {\cal B}_3 (t) \right\} \ ,
\label{A}
\\
\phi(t) & = & \omega_s t + \omega_s \ R \ \cos \beta \ \cos(\Omega t +
\eta_0 - \lambda) \ .
\label{phi}
\end{eqnarray}
In equation (\ref{phi}) $R$ is the distance of the guiding center of
the LISA array, $o$, from the Solar System Barycenter (SSB), ($\beta,
\lambda$) are the ecliptic latitude and longitude respectively of the
source location in the sky, $\Omega = 2 \pi/{\rm year}$, and
$\eta_0$ defines the position of the LISA guiding center in the
ecliptic plane at time $t = 0$. Note that the functions $c_k (t)$,
$d_k (t)$, and ${\cal B}_k (t)$ ($k = 2, 3$) do not depend on $x$. The
analytic expressions for $c_k (t)$, and $d_k (t)$ are the same as
those given in equations (46,47) of reference \cite{KTV04}, while the
functions ${\cal B}_k (t)$ ($k = 2, 3$) are equal to
\begin{equation}
{\cal B}_k (t) = (a^{(1)} + i \ a^{(3)}) \ u_k (t) + (a^{(2)} + i \ a^{(4)}) \ v_k
(t) \ .
\label{B}
\end{equation}
The coefficients ($a^{(1)}, a^{(2)}, a^{(3)}, a^{(4)}$) depend only on
the two independent amplitudes of the gravitational wave signal,
($h_+$, $h_\times$), the polarization angle, $\psi$, and an arbitrary
phase, $\phi_0$, that the signal has at time $t = 0$. Their analytic
expressions are given in equations (41--44) of reference \cite{KTV04},
while the functions $u_k (t)$, and $v_k (t)$ ($k=2, 3$) are given in
equations (27,28) in the same reference.

Since most of the gravitational wave energy radiated by the galactic
WD-WD binaries will be present in the lower part of the LISA
sensitivity frequency band, say between $10^{-4} - 10^{-3}$ Hz, it is
useful to provide an expression for the Taylor expansion of the $X$
response in the long-wavelength limit (LWL), i.e. when the wavelength
of the gravitational wave signal is much larger than the LISA
armlength ($x << 1$). The LWL expression will allow us to analytically
describe the general features of the white dwarfs background in the
$X$-combination, and derive computationally efficient algorithms for
numerically simulating the WD-WD background in the LISA data.

The nth-order truncation, $X^{\rm GW}_{(n)} (t)$, of the Taylor
expansion of $X^{\rm GW}(t)$ in power series of $x$ can be written in
the following form
\begin{equation}
X^{\rm GW}_{(n)} (t) = Re \sum_{k=0}^n A^{(k)} (t) \ x^{k+2} \ e^{-i \phi (t)}
\ ,
\label{Xn}
\end{equation}
where the first three functions of time $A^{(k)} (t), \ \ k \le 2$
are equal to
\begin{eqnarray}
A^{(0)}  & = &  4 \ [{\cal B}_2 - {\cal B}_3 ] \ ,
\nonumber
\\
A^{(1)}  & = & 4 i \ [(d_2+2) \ {\cal B}_2  - (d_3+2) {\cal B}_3 ] \ ,
\nonumber
\\
A^{(2)}  & = & [2{d_3}^2  + 8 d_3 + \frac{28}{3} + \frac{1}{6}
{c_3}^2] \ {\cal B}_3 -  [2{d_2}^2  + 8 d_2 + \frac{28}{3} + \frac{1}{6} {c_2}^2] \ {\cal B}_2 \ .
\label{An}
\end{eqnarray}
Note that the form we adopted for $X^{\rm GW} (t)$ (equation \ref{X})
makes the derivation of the functions $A^{(k)} (t)$ particularly easy
since the dependence on $x$ in $A (x,t)$ is now limited only to the
coefficients in front of the two functions ${\cal B}_2 (t)$ and ${\cal B}_3 (t)$
(see equation (\ref{A})).

Although it is generally believed that the lowest order
long-wavelength expansion of the $X$ combination, $X^{\rm GW}_{(0)}$,
is sufficiently accurate in representing a gravitational wave signal
in the low-part of the LISA frequency band, there has not been in the
literature any quantitative analysis of the error introduced by
relying on such a zero-order approximation.  Since any TDI combination
will contain a linear superposition of tens of millions of signals, it
is crucial to estimate such an error as a function of the order of the
approximation, $n$. In order to determine how many terms we needed to
use for a given signal angular frequency, $\omega_s$, we relied on
the following `` matching function'' \cite{ETKN05}
\begin{equation}
M(X^{\rm GW}, X^{\rm GW}_{(n)}) \equiv  \sqrt{\frac{\int_{0}^{T} {[X^{\rm GW} (t) - X^{\rm
    GW}_{(n)} (t)]}^2 dt}{\int_{0}^{T} {[X^{\rm GW} (t)]}^2 dt}} \ .
\label{M}
\end{equation}
Equation (\ref{M}) estimates the percent root-mean-squared error
implied by using the $n^{\rm th}$ order LWL approximation. With $T = 1
{\rm year}$ we have found \cite{ETKN05} that at $5 \times 10^{-4}$ Hz,
for instance, the zero-order LWL approximation ($n = 0$) of the $X$
combination shows an r.m.s.\ deviation from the exact response equal
to about $10$ percent.  As expected, this inaccuracy increases for
signals of higher frequencies, becoming equal to $40$ percent at $2
\times 10^{-3}$ Hz.  With $n=1$ the accuracy improves showing that the
$X^{\rm GW}_{(1)}$ response deviates from the exact one with an r.m.s.
error smaller than $10$ percent in the frequency band ($10^{-4} - 2
\times 10^{-3}$) Hz.  In our simulation we have actually implemented
the $n=2$ LWL expansion because it was possible and easy to do.

\section{White Dwarf binary population distribution}
\label{WD_POP}

The gravitational wave signal radiated by a WD-WD binary system
depends on eight parameters, ($\phi_o, \iota, \psi, D, \beta,
\lambda, {\cal M}_c, \omega_s$), which are the constant phase of
the signal
 ($\phi_o$) at the starting time of the observation, the inclination
angle ($\iota$) of the angular momentum of the binary system
relative to the line of sight, the polarization angle ($\psi$)
describing the orientation of the wave polarization axes, the
distance ($D$) to the binary, the angles ($\lambda, \beta$)
describing the location of the source in the sky relative to the
ecliptic plane, the chirp mass
 (${\cal M}_c$), and the angular frequency ($\omega_s$) in the source
reference frame respectively.  Since it can safely be assumed that
the chirp mass ${\cal M}_c$ and the angular frequency $\omega_s$
are independent of the source location \cite{NYP01} and of the
remaining angular parameters $\phi_o, \iota, \psi$, and because
there are no physical arguments for preferred values of the
constant phase $\phi_o$ and the orientation of the binary given by
the angles $\iota$ and $\psi$, it follows that the joint
probability distribution, $P(\phi_o, \iota, \psi, D, \beta,
\lambda, {\cal M}_c, \omega_s)$, can be rewritten in the following
form
\begin{equation}
P(\phi_o, \iota, \psi, D, \beta, \lambda, {\cal M}_c, \omega_s) =
P_1 (\phi_o) P_2 (\iota) P_3 (\psi) P_4 (D, \beta, \lambda)
P_5({\cal M}_c, \omega_s) \ .
\end{equation}
In the implementation of our simulation we have assumed the angles
$\phi_o$ and $\psi$ to be uniformly distributed in the interval
$[0, 2 \pi)$, and $\cos\iota$ uniformly distributed in the
interval $[-1,1]$. We further assumed the binary systems to be
randomly distributed in the Galactic disc according to the
following axially symmetric distribution ${\cal P}_4 (R, z)$ (see
\cite{NYP01} Eq. (5))
\begin{equation}
{\cal P}_4 (R,z) = \frac{e^{-R/H} \ sech^2(z/z_o)}{4 \pi z_o H^2}
\ , \label{eq:Galdis}
\end{equation}
where $(R,z)$ are cylindrical coordinates with origin at the
galactic center, $H = 2.5$ kpc, and $z_o = 200$ pc, and it is
proportional to $P_4 (D, \lambda, \beta)$ through the Jacobian of
the coordinate transformation.  Note that the position of the Sun
in this coordinate system is given by $R_{\odot} = 8.5 \ {\rm
kpc}$ and $z_{\odot} = -30 \ {\rm pc}$. We then generate the
positions of the sources from the distribution given by Eq.
(\ref{eq:Galdis}) and map them to their corresponding ecliptic
coordinates $(D, \beta,\lambda)$.

The physical properties of the WD-WD population (${\cal M}_c
\equiv {(m_1 m_2)}^{3/ 5}/{(m_1 + m_2)}^{1/5}$, with $m_1$, $m_2$
being the masses of the two stars, and $\omega_s = 2 \pi f_s = 4
\pi/P_{\rm
  orb}$) are taken from the binary population synthesis simulation
discussed in \cite{NYP04}. For details on this simulation we refer
the reader to \cite{NYP04}, and for earlier work to
\cite{HBW90,EIS87,HB90,HB00,BH97,NYP01}. The basic ingredient for
these simulations is an approximate binary evolution code. A
representation of the complete Galactic population of binaries is
produced by evolving a large (typically $10^6$) number of binaries
from their formation to the current time, where the distributions
of the masses and separations of the initial binaries are
estimated from the observed properties of local binaries.  This
initial-to-final parameter mapping is then convolved with an
estimate of the binary formation rate in the history of the Galaxy
to obtain the total Galactic population of binaries at the present
time.  From these the binaries of interest can then be selected.
In principle this technique is very powerful, although the results
can be limited by the limited knowledge we have on many aspects of
binary evolution. For WD-WD binaries, the situation is better than
for many other populations, since the observed population of WD-WD
binaries allows us to gauge the models (e.g. \cite{NYP01}).

We also include the population of semi-detached WD-WD binaries
(usually referred to as AM CVn systems) that are discussed in
detail in \cite{NYP04}.  In these binaries one white dwarf
transfers its outer layers onto a companion white dwarf. Due to
the redistribution of mass in the system, the orbital period of
these binaries increases in time, even though the angular momentum
of the binary orbit still decreases due to gravitational wave
losses. The formation of these systems is very uncertain, mainly
due to questions concerning the stability of the mass transfer
(e.g. \cite{MNS04}).

From the models of the Galactic population of the detached WD-WD
binaries and AM CVn systems two dimensional histograms were created,
giving the expected number of both WD-WD binaries and AM CVn systems
currently present in the Galaxy as function of the $log$ of the GW
radiation frequency, $f_s (= \omega_s/2 \pi)$ and chirp mass,
$\mathcal{M}_c$.  In the case of the detached WD-WD binaries, the
$(\log f_s, \mathcal{M}_c)$ space was defined over the set $\log f_s
\in [-6, -1]$, $\mathcal{M}_c \in (0, 1.5]$ and contained $50 \ \times
\ 30$ grid points, while in the case of the AM CVn systems the region
is intrinsically smaller, $\log f_s \in [-4, -1.5]$, $\mathcal{M}_c
\in (0, 1.2]$, containing only $25 \ \times \ 24$ grid points. In
order to generate values for these distributions within each grid
rectangle we have used an interpolation method \cite{ETKN05}.

\begin{figure}
  \includegraphics[width=10cm]{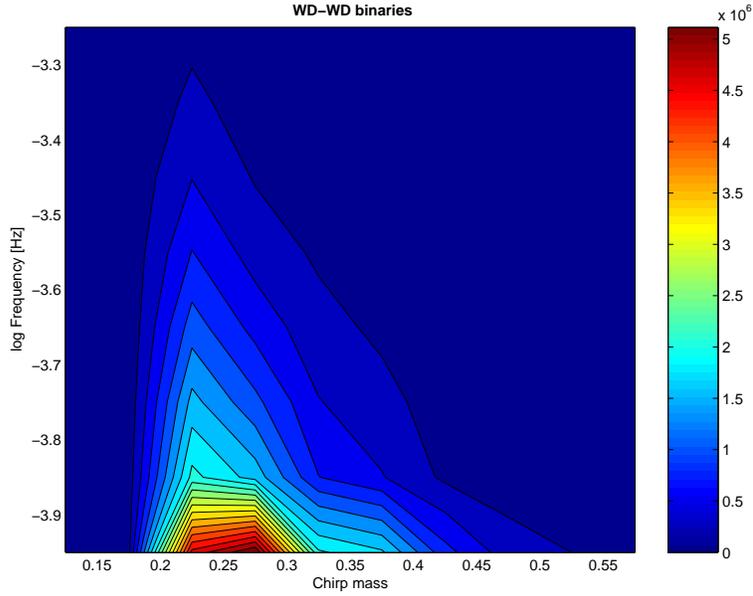}
  \caption{Distribution of detached white-dwarf --- white-dwarf
    binaries in our galaxy as a function of the gravitational wave
    frequency and chirp mass}
  \label{fig:wdwd_nel1}
\end{figure}

\begin{figure}
  \includegraphics[width=10cm]{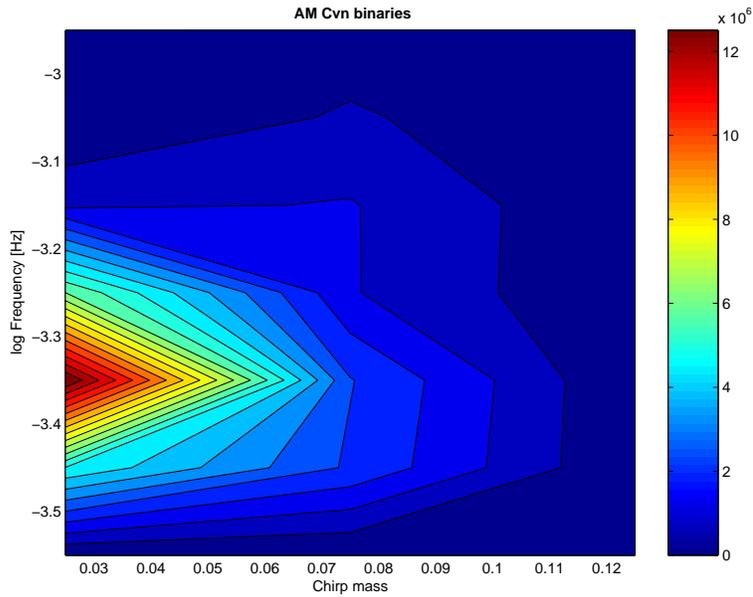}
  \caption{Distribution of  AM CVn binary systems in our galaxy as a
    function of the gravitational wave frequency and chirp  mass}
  \label{fig:amcvn_nel1}
\end{figure}

Figure (\ref{fig:wdwd_nel1}) shows the distribution of the number
of detached WD-WD binaries as a function of the signal frequency
and chirp mass in the form of a contour plot. This distribution
reaches its maximum within the LISA frequency band when the chirp
mass is equal to $\simeq 0.25 \ {\cal M}_\odot$, and it
monotonically decreases as a function of the signal frequency.
The distribution of the number of AM CVn systems has instead a
rather different shape, as shown by the contour plot given in
Figure (\ref{fig:amcvn_nel1}).


\section{Numerical Simulation}
\label{NumSim}

In order to simulate the LISA $X$ response to the population of WD-WD
binaries derived in Section \ref{WD_POP} one needs to coherently add
the LISA response to each individual signal. Although this could
naturally be done in the time domain, the actual CPU time required to
successfully perform such a simulation would be unacceptably long. The
generation in the time domain of one year of $X^{\rm GW} (t)$ response
to a single signal sampled at a rate of $16$ seconds would require
about $1$ second with an optimized C++ code running on a Pentium IV
$3.2$ GHz processor.  Since the number of signals from the background
is of the order $10^{8}$, it is clear that a different algorithm is
needed for simulating the background in the LISA data within a
reasonable amount of time. We were able to derive and implement
numerically an analytic formula of the Fourier transform of each
binary signal, which has allowed us to reduce the computational time
by almost a factor $100$.

\begin{figure}
\includegraphics[width=5in]{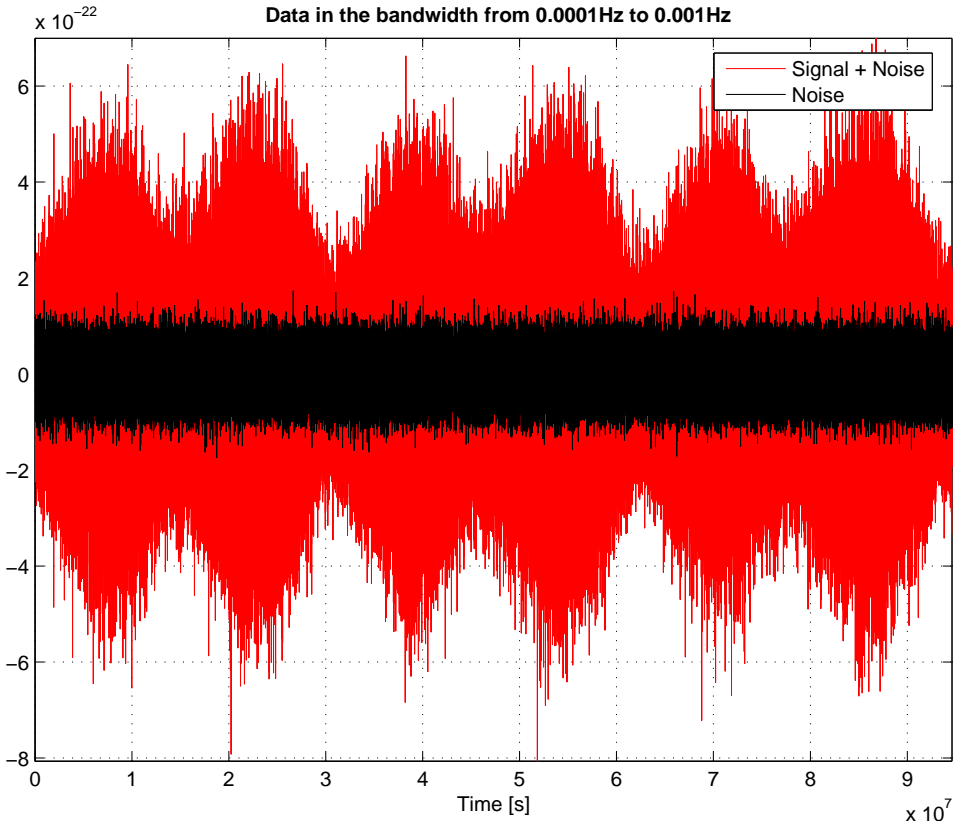}
\caption{Simulated time series of the WD-WD Galactic background signal
  of 3 years of data. The time series of LISA instrumental noise is
  displayed for comparison.}
\label{time_results}
\end{figure}

We have obtained an approximate analytic expression for the
finite-time Fourier transform of each WD-WD signal using the LWL
expansion (\ref{Xn}) with $n = 2$ and by applying the Nuttall's
modified Blackman-Harris window \cite{NBH81} in order to avoid
spectral leakage. We have then coherently added in the Fourier domain
all the signals radiated by the WD-WD galactic binary population
described in section (\ref{WD_POP}).  After inverse Fourier
transforming the synthesized response and then removing the window
from it, we finally obtained the time-domain representation of the
background as it will be seen in the LISA TDI combination $X$.  This
is shown in Figure (\ref{time_results}), where we plot three years
worth of simulated $X^{\rm GW} (t)$, and include the LISA noise
\cite{PPA98}. The one-year periodicity induced by the motion of LISA
around the Sun is clearly noticeable.  One other interesting feature
shown by Figure (\ref{time_results}) is that the amplitude response
reaches absolute minima when the Sun-LISA direction is roughly
oriented towards the Autumn equinox, while the absolute maxima take
place when the Sun-LISA direction is oriented roughly towards the
Galactic center \cite{Seto04}. This is because the ecliptic plane is
not parallel to the galactic plane, and our own solar system is about
$8.5 \ {\rm kpc}$ away from the galactic center (where most of the of
WD-WD binaries are concentrated). As a result it follows that the LISA
$X^{\rm GW}$ response to the WD-WD background does not have a
six-months periodicity.

Note also that, for a time period of about $2$ months, the absolute
minima reached by the amplitude of the LISA response to the WD-WD
background is only a factor less than $2$ larger than the level of the
instrumental noise.  This implies that during these observation times
LISA should be able to search for other sources of gravitational
radiation that are not located in the galactic plane. This might turn
out to be the easiest way to mitigate the detrimental effects of the
WD-WD background when searching for other sources of gravitational
radiation.  We will quantitatively analyze in a follow up work how to
take advantage of this observation in order to optimally search, during
these time periods, for sources that are off the galactic plane.

\section{Cyclostationary processes}
\label{sec:cyclo}

The results of our simulation (Figure \ref{time_results}) indicate
that the LISA $X^{\rm GW}$ response to the background can be regarded,
in a statistical sense, as a periodic function of time. This is
consequence of the deterministic (and periodic) motion of the LISA
array around the Sun. Since its autocorrelation function will also be
a periodic function of period one year, it follows that any LISA
response to the WD-WD background should no longer be treated as a
stationary random process but rather as a periodically correlated
random process. These kind of processes have been studied for many
years, and are usually referred to as cyclostationary random processes
(see \cite{H89} for a comprehensive overview of the subject and for more
references). In what follows we will briefly summarize the properties
of cyclostationary processes that are relevant to our problem.

A continuous stochastic process ${\cal X} (t)$ having finite second order
moments is said to be {\em cyclostationary\/} with period $T$ if the
following expectation values
\begin{eqnarray}
E[{\cal X} (t)] &=& m(t) = m(t + T) , \\
E[{\cal X} (t') {\cal X} (t)] &=& C(t',t) = C(t' + T,t + T)
\end{eqnarray}
are periodic functions of period $T$, for every $(t',t) \in {\bf R}
\times {\bf R}$. For simplicity from now on we will assume $m(t) = 0$.

If ${\cal X} (t)$ is cyclostationary, then the function $B(t,\tau) \equiv C(t
+ \tau,t)$ for a given $\tau \in {\bf R}$ is periodic with period $T$,
and it can be represented by the following Fourier series
\begin{equation}
B(t,\tau) = \sum_{r = -\infty}^{\infty}B_r(\tau) e^{i
2\pi\frac{r t}{T}} \ ,
\end{equation}
where the functions $B_r(\tau)$ are given by
\begin{equation}
B_r(\tau) = \frac{1}{T}\int^T_0B(t,\tau) e^{- i 2\pi r\frac{
    t}{T}} \ dt \ .
\label{eq:FB}
\end{equation}
The Fourier transforms $g_r(f)$ of $B_r(\tau)$ are the so called
``cyclic spectra'' of the cyclostationary process ${\cal X} (t)$  \cite{H89}
\begin{equation}
g_r(f) = \int_{-\infty}^{\infty}B_r(\tau)e^{-i 2\pi f \tau } \ d
\tau \ .
\label{eq:cykspec}
\end{equation}
If a cyclostationary process is real, the following relationships
between the cyclic spectra hold
\begin{eqnarray}
B_{-r}(\tau) & = & B^*_r(\tau) \ ,
\label{eq:ksym1}
\\
g_{-r}(-f) & = &  g^*_r(f) \ ,
\label{eq:ksym2}
\end{eqnarray}
where the symbol $^*$ means complex conjugation.  This implies that,
for a real cyclostationary process, the cyclic spectra with $r \geq 0$
contain all the information needed to characterize the process itself.

The function $\sigma^2(\tau) = B(0,\tau)$ is the variance of the
cyclostationary process ${\cal X} (t)$, and it can be written as a Fourier
decomposition as a consequence of Eq. (\ref{eq:FB})
\begin{equation}
\sigma^2(\tau) = \sum_{r=-\infty}^{\infty}H_r e^{ i
2\pi\frac{r \tau}{T}},
\label{eq:hars}
\end{equation}
where $H_r \equiv B_r(0)$ are harmonics of the variance $\sigma^2$.
From Eq. (\ref{eq:ksym1}) it follows that $H_{-r} = H^*_r$.

For a discrete, finite, real time series ${\cal X}_t$, $t = 1,\ldots,N$ we
can estimate the cyclic spectra by generalizing standard methods of
spectrum estimation used with stationary processes. Assuming again the
mean value of the time series ${\cal X}_t$ to be zero, the cyclic
autocorrelation sequences are defined as
\begin{equation}
s_l^r = \frac{1}{N}\sum_{t=1}^{N-|l|}{\cal X}_t {\cal X}_{t+|l|}
e^{-\frac{i 2\pi r (t-1)}{T}} \ .
\label{eq:cycorr}
\end{equation}
It has been shown \cite{H89} that the cyclic autocorrelations are
asymptotically (i.e.\ for $N \rightarrow \infty$) unbiased estimators
of the functions $B_r(\tau)$. The Fourier transforms of the cyclic
autocorrelation sequences $s_l^r$ are estimators of the cyclic spectra
$g_r(f)$. These estimators are asymptotically unbiased, and are called
``inconsistent estimators'' of the cyclic spectra, i.e. their
variances do not tend to zero asymptotically.  In the case of Gaussian
processes \cite{H89} consistent estimators can be obtained by first
applying a lag window to the cyclic autocorrelation and then perform
a Fourier transform.  This procedure represents a generalization of
the well-known technique for estimating the spectra of stationary
random processes \cite{PW93}.

An alternative procedure for identifying consistent estimators of the
cyclic spectra is to first take the Fourier transform,
$\tilde{{\cal X}}(f)$, of the time series ${\cal X} (t)$
\begin{equation}
\tilde{{\cal X}}(f) = \sum_{t = 1}^N  {\cal X}_t e^{-i 2\pi f (t - 1)}
\end{equation}
and then estimate the cyclic periodograms $g_r(f)$
\begin{equation}
g_r(f) = \frac{\tilde{{\cal X}}(f)\tilde{{\cal X}}^*(f - \frac{2\pi r}{T})}{N} \ .
\end{equation}
By finally smoothing the cyclic periodograms, consistent estimators of
the spectra $g_r(f)$ are then obtained.  The estimators of the
harmonics $H_r$ of the variance $\sigma^2$ of a cyclostationary random
process can be obtained by first forming a sample variance of the time
series ${\cal X}_t$. The sample variance is obtained by dividing the
time series ${\cal X}_t$ into contiguous segments of length $\tau_0$
such that $\tau_0$ is much smaller than the period $T$ of the
cyclostationary process, and by calculating the variance $\sigma^2_I$
over each segment.  Estimators of the harmonics can be obtained by
Fourier analyzing the series $\sigma^2_I$.  Note that the definitions
of (i) zero order ($r = 0$) cyclic autocorrelation, (ii) periodogram,
and (iii) zero order harmonic of the variance, coincide with those
usually adopted for stationary random processes.  Thus, even though a
cyclostationary time series is not stationary, ordinary spectral
analysis can be used for obtaining the zero order spectrum.  Note,
however, that cyclostationary random processes provide more spectral
information about the time series they are associated with due to the
existence of cyclic spectra with $r > 0$.

As an important and practical application, let us consider a time
series $y_t$ consisting of the sum of a stationary random process,
$n_t$, and a cyclostationary one ${\cal X}_t$ (i.e. $y_t = n_t + {\cal
  X}_t$).  Let the variance of the stationary time series be $\nu^2$
and its spectral density be $\mathcal{E}(f)$.  It is easy to see that
the resulting process is also cyclostationary. If the two processes
are uncorrelated, then the zero order harmonic $\Sigma^2_0$ of the
variance of the combined processes is equal to
\begin{equation}
\Sigma^2_0 = \nu^2 + \sigma^2_0 \ ,
\end{equation}
and the zero order spectrum $G_0(f)$ of $y_t$ is
\begin{equation}
G_0(f) =  \mathcal{E}(f)  +  g_0(f) \ .
\end{equation}
The harmonics of the variance as well as the cyclic spectra of $y_t$
with $r > 0$ coincide instead with those of ${\cal X}_t$.  In other words,
the harmonics of the variance and the cyclic spectra of the process
$y_t$ with $r > 0$ contain information only about the cyclostationary
process ${\cal X}_t$, and are not ``contaminated'' by the stationary process
$n_t$.

\section{Data analysis of the background signal}
\label{ANALYSIS}

\begin{figure}
\includegraphics[width=10cm]{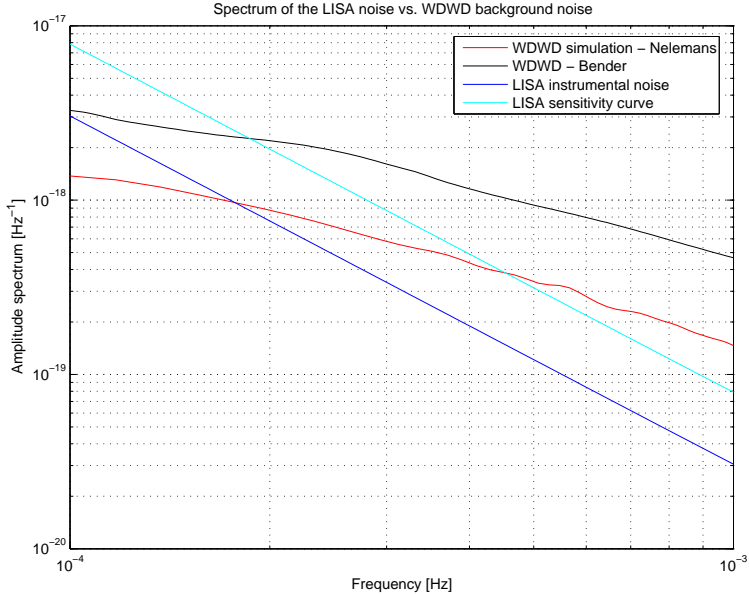}
\caption{Comparison of detached WD-WD background obtained from
binary population synthesis simulation ( \cite{NYP01,NYP04}) with
the WD-WD background calculated by Hils and Bender \cite{HB90}. Amplitude spectral
density of LISA instrumental noise and LISA sensitivity curve are drawn for comparison.
All spectral densities are one-sided.}
\label{fig:nel_ben_comp}
\end{figure}

We have numerically implemented the methods outlined in Section
\ref{sec:cyclo} and applied them to our simulated WD-WD background
signal.  A comparison of the results of our simulation of the detached
WD-WD background with the calculation of the background by Hils and
Bender \cite{HB90,Lweb} is shown in Figure \ref{fig:nel_ben_comp}.
Note that in Figure \ref{fig:nel_ben_comp} we have given the
spectra of the background that do not include LISA instrumental noise.
We find that the amplitude of the background from our simulation is a
factor of more than $2$ smaller than that of Hils and Bender.
The level of the WD-WD background is determined by the number of such
systems in the Galaxy. We estimate that our number WD-WD binaries should be
correct within a factor $5$ and thus the amplitude of the background should
be right within a factor of $\sqrt{5}$.
In Figure \ref{fig:nel_ben_comp} we have plotted the two backgrounds
against the LISA spectral density and we have also included the LISA
sensitivity curve. The latter is obtained by dividing the instrumental
noise spectral density by the detector GW transfer function averaged
over isotropically distributed and randomly polarized signals.  In the
zero-order long wavelength approximation this averaged transfer
function is equal to $\sqrt{3/20}$.  Note that for our simulation in
the region of the LISA band below $0.2$ millihertz the power of the
WD-WD background is smaller than that of the instrumental noise.

Our analysis was applied to $3$ years of LISA $X$ data consisting of a
coherent superposition of signals emitted by detached WD-WD binaries,
by semi-detached binaries (AM CVn systems), and of simulated
instrumental noise. The noise was numerically generated by using the
spectral density of the TDI $X$ observable given in \cite{ETA00}.  In
addition a $1$ mHz low-pass filter was applied to our data set in
order to focus our analysis to the frequency region in which the WD-WD
stochastic background is expected to be dominant.

\begin{figure}
\includegraphics[width=10cm]{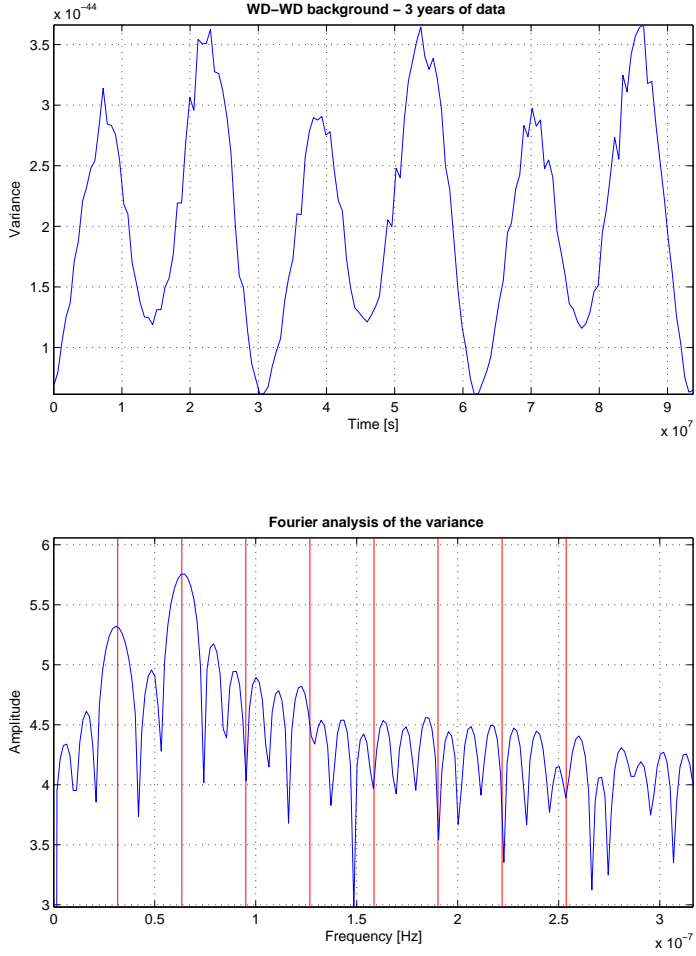}
\caption{Top panel: sample variance of the simulated WD-WD background.
  Three years of data are simulated. Data include two populations of
  the WD-WD binaries, detached and semi-detached ones added to the
  LISA instrumental noise. The data is passed through a low-pass
  filter with a cut-off frequency of 1 mHz.  Bottom panel: Fourier
  analysis of the sample variance.  Two harmonic are clearly
  resolved.\label{fig:wdwd_nel_var}}
\end{figure}

The results of the Fourier analysis of the sample variance of the
background signal are shown in Figure \ref{fig:wdwd_nel_var}. The
top panel of Figure \ref{fig:wdwd_nel_var} shows the sample
variance of the simulated data for which the variances were
estimated over a period of $1$ week; periodicity is clearly
visible.  The bottom panel instead shows the Fourier analysis of
the sample variance for which we have removed the mean from the
sample variance time series. The vertical lines correspond to
multiples of $1$ year; two harmonics can clearly be distinguished
from noise.  The other peaks of the spectrum that fall roughly
half way between the multiples of $1/{\rm year}$ frequency, are
from the rectangular window inherent to the finite time series.
In our analysis we have
plotted the absolute values $|H_r|$ of the coefficients of the
Fourier decomposition given by Eq.(\ref{eq:hars}). By using the
Fourier transform method, we cannot resolve higher than the second
harmonic in our $3$-year data set. We conclude that only the two
first harmonics can be extracted reliably from the data.

It is useful to compare the results of our numerical analysis against
the analytic calculations of Giampieri and Polnarev \cite{GP97}. Their
analytic expressions for the harmonics of the variance of a background
due to binary systems distributed in the galactic disc are given in
Eq. (42) and shown in Figure 4 of \cite{GP97}.  Our estimation roughly
matches theirs in that the 0th order harmonics is dominant and the
first two harmonics have more power than the remaining ones. Our
estimate of the power in the second harmonic, however, is larger than
that in the first one, whereas they find the opposite.  We attribute
this difference to their use of a Gaussian distribution of sources in
the Galactic disc rather than the (usually assumed) exponential that
we adopted.  Comparison between these two results suggests that it
should be possible to infer the distribution of WD-WD binaries in our
Galaxy by properly analyzing the harmonics of the variance of the
galactic background measured by LISA.  How this can be done will be
the subject of a future work.

\begin{figure}
\includegraphics[width=10cm]{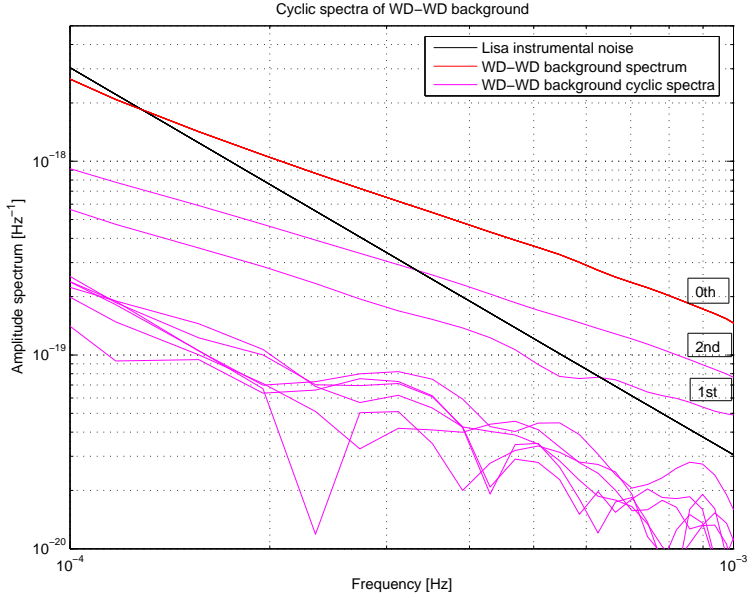}
\caption{The main (k = 0) spectrum of simulated WD-WD background
signal (red). 8 cyclic spectra (magenta) estimated from the
simulated data. Spectral density of LISA instrumental noise
(black) is shown for comparison. The $0$th order spectrum contains
LISA instrumental and hence it differs from the spectrum given in
Figure \ref{fig:nel_ben_comp}. \label{fig:spec_cyclo}}
\end{figure}

As a next step in our analysis, we have estimated the cyclic
spectra $g_r(f)$ (Eq.(\ref{eq:cykspec} with $r = 1, 2, ... 8$) of
the simulated WD-WD background signal. The number $8$ comes from
the theoretical predictions of the number of
harmonics. The estimated absolute values of the cyclic spectra
are shown in Figure \ref{fig:spec_cyclo}, where we also
plot the spectrum of the LISA instrumental noise and the main
spectrum ($r = 0$) estimated from the simulation. We find that the
main spectrum and the first two cyclic spectra $r = 1$ and $r = 2$
have the largest magnitude and, over some frequency range, lie
above the LISA instrumental noise. The remaining spectra are an
order of magnitude smaller and are very noisy. We may also notice
that all the cyclic spectra have roughly the same slope. This
follows from the assumed independence between the location of the
binaries in the Galaxy ($D, \lambda, \beta$) and their frequencies
and chirp masses (${\cal M}_c, \omega_s$). We also find the
magnitude of the 2nd cyclic spectrum to be higher than the first,
similarly to what we had for the harmonics of the variance.
Note that we estimated the spectra from the time series consisting
of the WD-WD background added to the LISA instrumental noise.

Our analysis has shown that the LISA data will allow us to compute
$17$ independent cyclic spectra (the $8$ complex cyclic spectra
$g_r(f) , r=1, 2, ...8$ and the real spectrum $g_0 (f)$) of the WD-WD
galactic background, $5$ of which can be expected to be measured
reliably.  We have also shown that by performing generalized spectral
analysis of the LISA data we will be able to derive more information
about the WD-WD binary population (properties of the distribution of
its parameters) than we would have by only looking at the ordinary
$g_0 (f)$ spectrum.

\section*{Acknowledgments}
The supercomputers used in this investigation were provided by funding
from the Jet Propulsion Laboratory Institutional Computing and
Information Services, and the NASA Directorates of Aeronautics
Research, Science, Exploration Systems, and Space Operations.  A.K.\
acknowledges support from the National Research Council under the
Resident Research Associateship program at the Jet Propulsion
Laboratory. This work was supported in part by Polish Science
Committee Grant No. KBN 1 P03B 029 27.  This research was performed at
the Jet Propulsion Laboratory, California Institute of Technology,
under contract with the National Aeronautics and Space Administration.

\end{document}